\documentclass[preprint]{elsarticle}

\usepackage{lineno,hyperref}
\usepackage{graphicx,amsmath,amssymb,amsthm,bm}
\modulolinenumbers[5]

\journal{}

\newtheorem{definition}{Definition}

\newtheorem{lemma}{Lemma}
\newtheorem{theorem}{Theorem}

\newtheorem{assumption}{Assumption}

\begin{document}

\begin{frontmatter}

\title{Feedback autocratic strategies in stochastic games}

\author{Masahiko Ueda\corref{mycorrespondingauthor}}
\ead{ueda@tmu.ac.jp}
\address{Department of Physics, Tokyo Metropolitan University, Tokyo 192-0397, Japan}

\begin{abstract}
A stochastic game is an optimal control problem for multiple agents in a stochastic environment.
Stochastic games contain important practical problems, related to environmental issues or traffic flows.
If the number of environmental states is one, such stochastic games are called repeated games.
In repeated games, several strategies which unilaterally control payoffs have been proposed, and these strategies work in a mechanism similar to feedback control.
In this paper, we generalize these strategies as feedback strategies in stochastic games, and investigate their properties.
In particular, we provide a necessary and sufficient condition for such feedback strategies to work as controllers of payoffs under some assumptions.
We demonstrate the performance of such feedback strategies in several examples related to social dilemmas.
Our findings connect the theory of payoff control in repeated games to the theory of feedback control in stochastic games.
\end{abstract}


\end{frontmatter}


\section{Introduction}
\label{sec:introduction}
Feedback control is the central concept of classical control theory \cite{Wie1948}.
When an output increases, the feedback system decreases the output.
Conversely, when an output decreases, the feedback system increases the output.
For linear dynamical systems modeled by ordinary differential equations, the stability of feedback control has been well understood, by methods as the Nyquist stability criterion \cite{DFT2013}.
Even in optimal control problems, feedback control is derived as an optimal regulator for linear-quadratic systems \cite{KwaSiv1972}.
The concept of feedback control also appears in systems biology \cite{Alo2019} and economics \cite{Art2018}, not limited to control theory.

In nonlinear systems, control theory is mainly described in the framework of optimal control problems \cite{Kir2004}.
An optimal control problem is one in which an agent finds the control minimizing a cost function for a dynamical system.
The theory of optimal control was further extended to two directions.
In the first direction, optimal control problems with multiple agents were introduced, and these problems are called differential games \cite{Isa1999}.
In the second direction, optimal control problems for stochastic systems were introduced, and these problems are called Markov decision processes \cite{SutBar2018}.
Furthermore, a simultaneous extension of optimal control problems in these two directions is a stochastic game \cite{Sha1953,Sol2022}.
In a stochastic game, each agent optimizes a different objective function in a stochastic environment.
Stochastic games contain important practical problems, related to environmental issues \cite{HSCN2018} or traffic flows \cite{HuWel2003}.
Many numerical algorithms for stochastic games have also been proposed \cite{SPG2007,BBD2008}.

If the number of environmental states is one, such stochastic games are called repeated games \cite{FudTir1991}.
In a repeated game, each agent optimizes a different objective function (payoff) in a static environment.
In repeated games, a class of strategies called zero-determinant strategies can be used to control payoffs \cite{PreDys2012,McAHau2016,HCN2018}.
Generally, zero-determinant strategies unilaterally enforce linear relationships between payoffs.
In the repeated prisoner's dilemma game, zero-determinant strategies contain ones setting the opponent's payoff \cite{BNS1997}, unbeatable strategies \cite{PreDys2012}, and ones promoting mutual cooperation \cite{StePlo2013}.
Furthermore, in the repeated prisoner's dilemma game, an optimal control strategy called the Grim Trigger strategy was interpreted in terms of control of conditional expectations of payoffs \cite{Ued2022c}.
It was also pointed out that zero-determinant strategies work in a mechanism similar to feedback control \cite{Ued2025}.
Recently, the performance of zero-determinant strategies has been investigated in stochastic games \cite{DRWW2021,LiuWu2022,MMHet2025,WZWL2026,ZYRW2026,NakUed2027}.
In this manner, the theory of payoff control in repeated games and control theory are approaching each other.

Although stochastic games have been analyzed as optimal control problems, the performance of feedback control in stochastic games itself rather remains unclear.
In the repeated prisoner's dilemma game, a strategy controlling the opponent's payoff more directly than zero-determinant strategies was discussed \cite{BNS1997}.
Furthermore, a strategy of cumulative reciprocity, which unilaterally equalizes the payoffs of two players in the repeated prisoner's dilemma game, was recently introduced \cite{LZLet2022}.
These strategies can be regarded as feedback control, and it is expected that we can generalize these strategies even in stochastic games.

In this paper, we introduce the concept of feedback strategies in stochastic games.
These strategies simply decrease (increase) the output when the output increases (decreases).
We then investigate when these feedback strategies can be used to control payoffs unilaterally.
Through several examples, we find that these feedback strategies can be implemented more easily than zero-determinant strategies.

\section{Preliminaries}
\label{sec:preliminaries}
\subsection{Stochastic games}
A stochastic game is given by $G:=\left( \mathcal{N}, \Sigma, \left\{ A_j \right\}_{j\in \mathcal{N}}, T_\mathrm{E}, P_\mathrm{E}^{(1)}, \left\{ s_j \right\}_{j\in \mathcal{N}} \right)$ \cite{Sha1953,HSCN2018}.
$\mathcal{N}$ is the set of players.
$\Sigma$ is the set of states of an environment.
$A_j$ is the set of actions of player $j$.
$T_\mathrm{E}: \prod_{k\in \mathcal{N}}A_k \times \Sigma \rightarrow \Delta(\Sigma)$ is the transition probability of states, where $\Delta(\Sigma)$ is the probability simplex on $\Sigma$.
$P_\mathrm{E}^{(1)}$ is the probability distribution of the initial state.
$s_j: \prod_{k\in \mathcal{N}}A_k \times \Sigma \rightarrow \mathbb{R}$ is the one-shot payoff function of player $j$.
We collectively write $\mathcal{A}:= \prod_{k\in \mathcal{N}} A_k$ and $A_{-j}:= \prod_{k\neq j} A_k$ for all $j \in \mathcal{N}$.
We also introduce the notations $\bm{a}:=\left( a_k \right)_{k\in \mathcal{N}} \in \mathcal{A}$ and $a_{-j}:=\left( a_k \right)_{k\neq j} \in A_{-j}$.
The players can choose actions in each round referring to all histories of actions and states, and we call such plans \emph{strategies}.
When we write an action profile and a state at the $t$-th round as $\bm{a}^{(t)}$ and $\sigma^{(t)}$ with $t\geq 1$, respectively, the (undiscounted) payoff of player $j$ in the stochastic game is defined by
\begin{align}
 \mathcal{S}_j &:= \lim_{T\rightarrow \infty} \frac{1}{T} \sum_{t=1}^T \mathbb{E} \left[ s_j\left( \bm{a}^{(t)}, \sigma^{(t)} \right) \right],
\end{align}
where $\mathbb{E}[\cdot]$ is the expectation with respect to strategies of all players and the transition probability $T_\mathrm{E}$.
If such limit does not exist, we replace $\lim$ by $\liminf$.

In this paper, we assume that $\Sigma$ and $A_k$ $(\forall k \in \mathcal{N})$ are finite sets.

\subsection{Feedback autocratic strategies}
The purpose of this paper is constructing strategies which unilaterally control the quantity $B:\mathcal{A}\times \Sigma \rightarrow \mathbb{R}$ in the following sense:
\begin{definition}
A strategy of player $j$ is an \emph{autocratic} strategy controlling $B$ if it controls the quantity $B:\mathcal{A}\times \Sigma \rightarrow \mathbb{R}$ as
\begin{align}
 \left| \sum_{t=1}^T \mathbb{E} \left[ B\left( \bm{a}^{(t)}, \sigma^{(t)} \right) \right] \right| &\leq \Delta \quad (\forall T)
 \label{eq:autocratic}
\end{align}
with some constant $\Delta\geq 0$ regardless of the strategies of the other players and for arbitrary initial conditions of player $j$ and the environmental state.
\end{definition}
Typically, $B$ is related to payoffs $\left\{ s_k \right\}_{k\in \mathcal{N}}$.
In repeated games, which are stochastic games with one environmental state, zero-determinant strategies are autocratic strategies in this sense \cite{Ued2026sub}.

In this paper, we make two assumptions.
\begin{assumption}
\label{asp:deterministic}
The transition probability $T_\mathrm{E}$ is deterministic: There exists a transition function $f_\mathrm{E}: \mathcal{A}\times \Sigma \rightarrow \Sigma$ such that
\begin{align}
 \sigma^{(t+1)} &= f_\mathrm{E}\left( \bm{a}^{(t)}, \sigma^{(t)} \right) \quad (\forall t\geq 1).
\end{align}
\end{assumption}
\begin{assumption}
\label{asp:singleton}
For every $\sigma\in \Sigma$, both
\begin{align}
 \arg\max_{a_j} \left[ \min_{a_{-j}} B\left( \bm{a}, \sigma \right) \right]
\end{align}
and
\begin{align}
 \arg\min_{a_j} \left[ \max_{a_{-j}} B\left( \bm{a}, \sigma \right) \right]
\end{align}
are singletons.
\end{assumption}

Due to Assumption \ref{asp:singleton}, we can define the unique actions
\begin{align}
 \overline{a}_j\left( \sigma \right) &:= \arg\max_{a_j} \left[ \min_{a_{-j}} B\left( \bm{a}, \sigma \right) \right]
 \label{eq:aj_ov} \\
 \underline{a}_j\left( \sigma \right) &:= \arg\min_{a_j} \left[ \max_{a_{-j}} B\left( \bm{a}, \sigma \right) \right].
 \label{eq:aj_und}
\end{align}

We now introduce the feedback strategies as follows:
\begin{definition}
The \emph{feedback} strategy of player $j$ about $B$ is such that
\begin{align}
 a_j^{\mathrm{F}(t)} &:= \left\{
  \begin{array}{ll}
    \overline{a}_j\left( \sigma^{(t)} \right) & \left( \mathrm{if} \quad \sum_{t^\prime=1}^{t-1} B\left( \bm{a}^{(t^\prime)}, \sigma^{(t^\prime)} \right) \leq 0 \right) \\
    \underline{a}_j\left( \sigma^{(t)} \right) & \left( \mathrm{if} \quad \sum_{t^\prime=1}^{t-1} B\left( \bm{a}^{(t^\prime)}, \sigma^{(t^\prime)} \right) > 0 \right)
  \end{array}
  \right.
  \label{eq:feedback}
\end{align}
with some initial condition.
\end{definition}
The idea behind feedback strategies is that they simply take actions which are expected to increase (decrease) total $B$ when total $B$ is negative (positive).
If a feedback strategy about $B$ is autocratic, we call it a \emph{feedback autocratic strategy}.
An example of feedback autocratic strategies was originally discussed in the context of equalizer strategies in the repeated prisoner's dilemma game \cite{BNS1997}, which unilaterally control the payoff of the opponent.
Later, a strategy of cumulative reciprocity was introduced in the repeated prisoner's dilemma game \cite{LZLet2022}, which can be regarded as a feedback autocratic strategy controlling the payoff difference between two players.

\section{Results}
\label{sec:results}
\subsection{Existence condition}
\label{subsec:existence}
Here, we consider the situation where player $j$ adopts the feedback strategy about $B$.
Under Assumptions \ref{asp:deterministic} and \ref{asp:singleton}, a stochastic game behaves as a couple of deterministic finite automata where the set of states is $\Sigma$ and the set of input symbols is $A_{-j}$.
The transition functions of two finite automata are $f_\mathrm{E}$ with $a_j=\overline{a}_j\left( \sigma \right)$ and $a_j=\underline{a}_j\left( \sigma \right)$, respectively.
The dynamics of the state in each automaton is described on a directed graph where states correspond to vertices and edges are connected with input symbols.
The two automata switch to each other according to $a_j^{\mathrm{F}(t)}=\overline{a}_j\left( \sigma^{(t)} \right)$ or $a_j^{\mathrm{F}(t)}=\underline{a}_j\left( \sigma^{(t)} \right)$.
We call $\overline{a}_j$ and $\underline{a}_j$ \emph{modes} of the automaton.

Now, we prove the following theorem on the existence of feedback autocratic strategies.
\begin{theorem}
\label{thm:existence}
A feedback strategy of player $j$ about $B$ is autocratic if and only if the inequality
\begin{align}
 \sum_{t=1}^{T_C} B\left( \overline{a}_j\left( \sigma^{(t)} \right), a_{-j}^{(t)}, \sigma^{(t)} \right) &\geq 0
 \label{eq:condition_ov}
\end{align}
holds for each possible cycle $C$ (with period $T_C$) in mode $\overline{a}_j$, and the inequality
\begin{align}
 \sum_{t=1}^{T_C} B\left( \underline{a}_j\left( \sigma^{(t)} \right), a_{-j}^{(t)}, \sigma^{(t)} \right) &\leq 0
 \label{eq:condition_und}
\end{align}
holds for each possible cycle $C$ (with period $T_C$) in mode $\underline{a}_j$.
\end{theorem}

\begin{proof}
(Necessity.)
Suppose that there exists a cycle $C$ such that
\begin{align}
 \sum_{t=1}^{T_C} B\left( \overline{a}_j\left( \sigma^{(t)} \right), a_{-j}^{(t)}, \sigma^{(t)} \right) &< 0
\end{align}
or
\begin{align}
 \sum_{t=1}^{T_C} B\left( \underline{a}_j\left( \sigma^{(t)} \right), a_{-j}^{(t)}, \sigma^{(t)} \right) &> 0
\end{align}
hold.
Then, according to Lemma \ref{lem:cycle} (in \ref{app:lemma}), there exists a trajectory in which cycle $C$ is infinitely repeated without switching the mode.
If such a cycle $C$ is infinitely repeated, it violates Eq. (\ref{eq:autocratic}).
Therefore, the feedback strategy cannot be autocratic.

(Sufficiency.)
Suppose that the conditions (\ref{eq:condition_ov}) and (\ref{eq:condition_und}) hold for all possible cycles.
Without loss of generality, we consider the situation where player $j$ starts with mode $\overline{a}_j$.
The realized trajectory is such that $\left( a_j^{(1)}, a_{-j}^{(1)}, \sigma^{(1)} \right) \rightarrow \left( \overline{a}_j\left( \sigma^{(2)} \right), a_{-j}^{(2)}, \sigma^{(2)} \right) \rightarrow \cdots$.
In mode $\overline{a}_j$, the states are classified into transient states, which cannot be visited infinitely many times, and non-transient states.
It should be noted that all cycles in mode $\overline{a}_j$ are realized on non-transient states in mode $\overline{a}_j$.
If the inequality $\sum_{t^\prime=1}^{t} B\left( \bm{a}^{(t^\prime)}, \sigma^{(t^\prime)} \right) > 0$ is satisfied while the state is transient, the mode switches to $\underline{a}_j$.
Otherwise, the state reaches non-transient ones during mode $\overline{a}_j$.
Because the number of non-transient states is finite, at least one state can be visited infinitely many times if the mode is $\overline{a}_j$ forever.
In the sequence $\left\{ \sigma^{(t)} \right\}_{t=1}^{\infty}$ for the case where the mode is $\overline{a}_j$ forever, we search the first $\tau$ such that $\sigma^{(\tau)}$ coincides with $\sigma^{(t^\prime)}$ with $1\leq t^\prime \leq \tau-1$.
This constitutes a cycle, and, according to Eq. (\ref{eq:condition_ov}), we observe
\begin{align}
 \sum_{t=t^\prime}^{\tau-1} B\left( \overline{a}_j\left( \sigma^{(t)} \right), a_{-j}^{(t)}, \sigma^{(t)} \right) &\geq 0.
\end{align}
Then, we remove the part $\left\{ \sigma^{(t)} \right\}_{t=t^\prime}^{\tau-1}$ from $\left\{ \sigma^{(t)} \right\}_{t=1}^{\infty}$, and repeat this procedure.
By this, we can extract infinitely many non-negative quantities from $\sum_{t=1}^{\infty} B\left( \overline{a}_j\left( \sigma^{(t)} \right), a_{-j}^{(t)}, \sigma^{(t)} \right)$.
This leads to $\sum_{t^\prime=1}^{t} B\left( \bm{a}^{(t^\prime)}, \sigma^{(t^\prime)} \right) > 0$ with some $t$ or $\sum_{t^\prime=1}^{t} B\left( \bm{a}^{(t^\prime)}, \sigma^{(t^\prime)} \right) \leq 0$ for all $t\geq 1$.
In the former case, the mode switches to $\underline{a}_j$.
In the latter case, all parts removed infinitely many times satisfy
\begin{align}
 \sum_{t=t^\prime}^{\tau-1} B\left( \overline{a}_j\left( \sigma^{(t)} \right), a_{-j}^{(t)}, \sigma^{(t)} \right) &= 0.
\end{align}
For such a case, $\sum_{t=1}^{\infty} B\left( \overline{a}_j\left( \sigma^{(t)} \right), a_{-j}^{(t)}, \sigma^{(t)} \right)$ remains finite.

If player $j$ starts with mode $\underline{a}_j$, the mode also eventually switches to $\overline{a}_j$ or $\sum_{t=1}^{\infty} B\left( \underline{a}_j\left( \sigma^{(t)} \right), a_{-j}^{(t)}, \sigma^{(t)} \right)$ remains finite.

According to the consideration above, we can estimate $\Delta$ in Eq. (\ref{eq:autocratic}).
We define $\beta := \max_{\bm{a},\sigma} \left| B(\bm{a}, \sigma) \right|$.
We write the number of transient states as $N_\mathrm{tr}$ and the number of non-transient states as $N_\mathrm{ntr}$ in each mode.
It should be noted that $N_\mathrm{tr} + N_\mathrm{ntr} = \left| \Sigma \right|$.
The number of each non-transient state in one trajectory before switching of modes is at most one after removal of cycles.
Therefore, we obtain $\Delta = \beta \left( N_\mathrm{tr} + N_\mathrm{ntr} + 1 \right) = \beta \left( \left| \Sigma \right| + 1 \right)$, where $1$ comes from initial accumulation after switching.
\end{proof}

It should be noted that we have not used the concrete definitions (\ref{eq:aj_ov}) and (\ref{eq:aj_und}) in the proof of Theorem \ref{thm:existence}.
Therefore, Theorem \ref{thm:existence} itself holds for arbitrary choice of $\overline{a}_j$ and $\underline{a}_j$ in Eq. (\ref{eq:feedback}).
However, for bad choice of $\overline{a}_j$ and $\underline{a}_j$, satisfying the conditions (\ref{eq:condition_ov}) and (\ref{eq:condition_und}) is expected to become impossible.

The following theorem gives a sufficient condition for the existence of feedback autocratic strategies of player $j$ controlling $B$.
\begin{theorem}
\label{thm:existence_sufficient}
If the inequalities
\begin{align}
 B\left( \overline{a}_j\left( \sigma \right), a_{-j}, \sigma \right) &\geq 0 \quad \left( \forall a_{-j} \right) \nonumber \\
 B\left( \underline{a}_j\left( \sigma \right), a_{-j}, \sigma \right) &\leq 0 \quad \left( \forall a_{-j} \right)
 \label{eq:existence_sufficient}
\end{align}
hold for all $\sigma$, then a feedback strategy of player $j$ about $B$ is autocratic.
\end{theorem}

\begin{proof}
If the inequalities (\ref{eq:existence_sufficient}) hold for all $\sigma$, the conditions (\ref{eq:condition_ov}) and (\ref{eq:condition_und}) are trivially satisfied for all possible cycles.
\end{proof}

We remark that the condition (\ref{eq:existence_sufficient}) is equivalent to
\begin{align}
 \max_{a_j} \min_{a_{-j}} B\left( a_j, a_{-j}, \sigma \right) &\geq 0 \nonumber \\
 \min_{a_j} \max_{a_{-j}} B\left( a_j, a_{-j}, \sigma \right) &\leq 0
\end{align}
for all $\sigma$.
This condition is also equivalent to the existence condition of zero-determinant strategies \cite{Ued2022b} when we consider each state independently.

\subsection{Periodic games}
\label{subsec:periodic}
As a special class of stochastic games, a \emph{periodic game} is one where $L:= \left| \Sigma \right|$ states are periodically realized regardless of the actions of players \cite{MMHet2025,NakUed2027}.
We define $\Sigma = \left\{ \sigma_1, \cdots, \sigma_L \right\}$, and write the order of states as $\sigma_1 \rightarrow \sigma_2 \rightarrow \cdots \rightarrow \sigma_L$.
For periodic games, Theorem \ref{thm:existence} is modified as follows.

\begin{theorem}
\label{thm:existence_periodic}
For periodic games, a feedback strategy of player $j$ about $B$ is autocratic if and only if the inequalities
\begin{align}
 \sum_{t=1}^{L} B\left( \overline{a}_j\left( \sigma_t \right), a_{-j}^{(t)}, \sigma_t \right) &\geq 0 \quad \left( \forall \left\{ a_{-j}^{(t)} \right\}_{t=1}^L \right)
 \label{eq:condition_ov_per}
\end{align}
and
\begin{align}
 \sum_{t=1}^{L} B\left( \underline{a}_j\left( \sigma_t \right), a_{-j}^{(t)}, \sigma_t \right) &\leq 0 \quad \left( \forall \left\{ a_{-j}^{(t)} \right\}_{t=1}^L \right)
 \label{eq:condition_und_per}
\end{align}
hold.
\end{theorem}

\begin{proof}
For periodic games with period $L$, the only cycle of states is $\sigma_1 \rightarrow \sigma_2 \rightarrow \cdots \rightarrow \sigma_L$.
In this cycle, each $a_{-j}^{(t)}$ is arbitrary.
Therefore, the conditions (\ref{eq:condition_ov}) and (\ref{eq:condition_und}) are reduced to the conditions (\ref{eq:condition_ov_per}) and (\ref{eq:condition_und_per}), respectively.
\end{proof}

We remark that the conditions (\ref{eq:condition_ov_per}) and (\ref{eq:condition_und_per}) are equivalent to
\begin{align}
 \sum_{t=1}^{L} \max_{a_j} \min_{a_{-j}} B\left( a_j, a_{-j}, \sigma_t \right) &\geq 0 \nonumber \\
 \sum_{t=1}^{L} \min_{a_j} \max_{a_{-j}} B\left( a_j, a_{-j}, \sigma_t \right) &\leq 0. 
 \label{eq:existence_multichannel}
\end{align}
This condition is equivalent to the existence condition of zero-determinant strategies in multichannel games \cite{Ued2026}, where multiple repeated games are played in parallel \cite{DHNH2020}.
The equivalence between the existence condition of payoff-controlling strategies in multichannel games and that in periodic games was conjectured in Ref. \cite{NakUed2027}.

\subsection{Stochastic prisoner's dilemma}
\label{subsec:SPD}
As an example, we consider a stochastic prisoner's dilemma game \cite{HSCN2018}.
In this game, $\mathcal{N}= \{ 1, 2 \}$, $\Sigma= \{ \sigma_1, \sigma_2 \}$, $A_j= \left\{ C, D \right\}$ $(j=1, 2)$, and the payoffs $\left\{ s_j \right\}_{j\in \mathcal{N}}$ are given as in Tables \ref{table:PD1_Hilbe} and \ref{table:PD2_Hilbe}, where $T_n>R_n>P_n>S_n$ and $2R_n>T_n+S_n$ $(n=1, 2)$ are assumed.
The actions $C$ and $D$ represent cooperation and defection, respectively.
The transition function $f_\mathrm{E}$ is given as
\begin{align}
 f_\mathrm{E}\left( C, C, \sigma_1 \right) &= \sigma_1 \nonumber \\
 f_\mathrm{E}\left( C, D, \sigma_1 \right) &= \sigma_2 \nonumber \\
 f_\mathrm{E}\left( D, C, \sigma_1 \right) &= \sigma_2 \nonumber \\
 f_\mathrm{E}\left( D, D, \sigma_1 \right) &= \sigma_2 \nonumber \\
 f_\mathrm{E}\left( C, C, \sigma_2 \right) &= \sigma_1 \nonumber \\
 f_\mathrm{E}\left( C, D, \sigma_2 \right) &= \sigma_2 \nonumber \\
 f_\mathrm{E}\left( D, C, \sigma_2 \right) &= \sigma_2 \nonumber \\
 f_\mathrm{E}\left( D, D, \sigma_2 \right) &= \sigma_2.
\end{align}
The states $\sigma_1$ and $\sigma_2$ correspond to a good state and a bad state, respectively, and transition to the bad state is coupled to defection of some players.
\begin{table}[tb]
  \centering
  \caption{Payoffs in state $\sigma_1$.}
  \begin{tabular}{|c|cc|} \hline
    & $C$ & $D$ \\ \hline
   $C$ & $R_1, R_1$ & $S_1, T_1$ \\
   $D$ & $T_1, S_1$ & $P_1, P_1$ \\ \hline
  \end{tabular}
  \label{table:PD1_Hilbe}
  
  \centering
  \caption{Payoffs in state $\sigma_2$.}
  \begin{tabular}{|c|cc|} \hline
    & $C$ & $D$ \\ \hline
   $C$ & $R_2, R_2$ & $S_2, T_2$ \\
   $D$ & $T_2, S_2$ & $P_2, P_2$ \\ \hline
  \end{tabular}
  \label{table:PD2_Hilbe}
\end{table}

\begin{figure}[tbp]
\begin{center}
\includegraphics[clip, width=8.0cm]{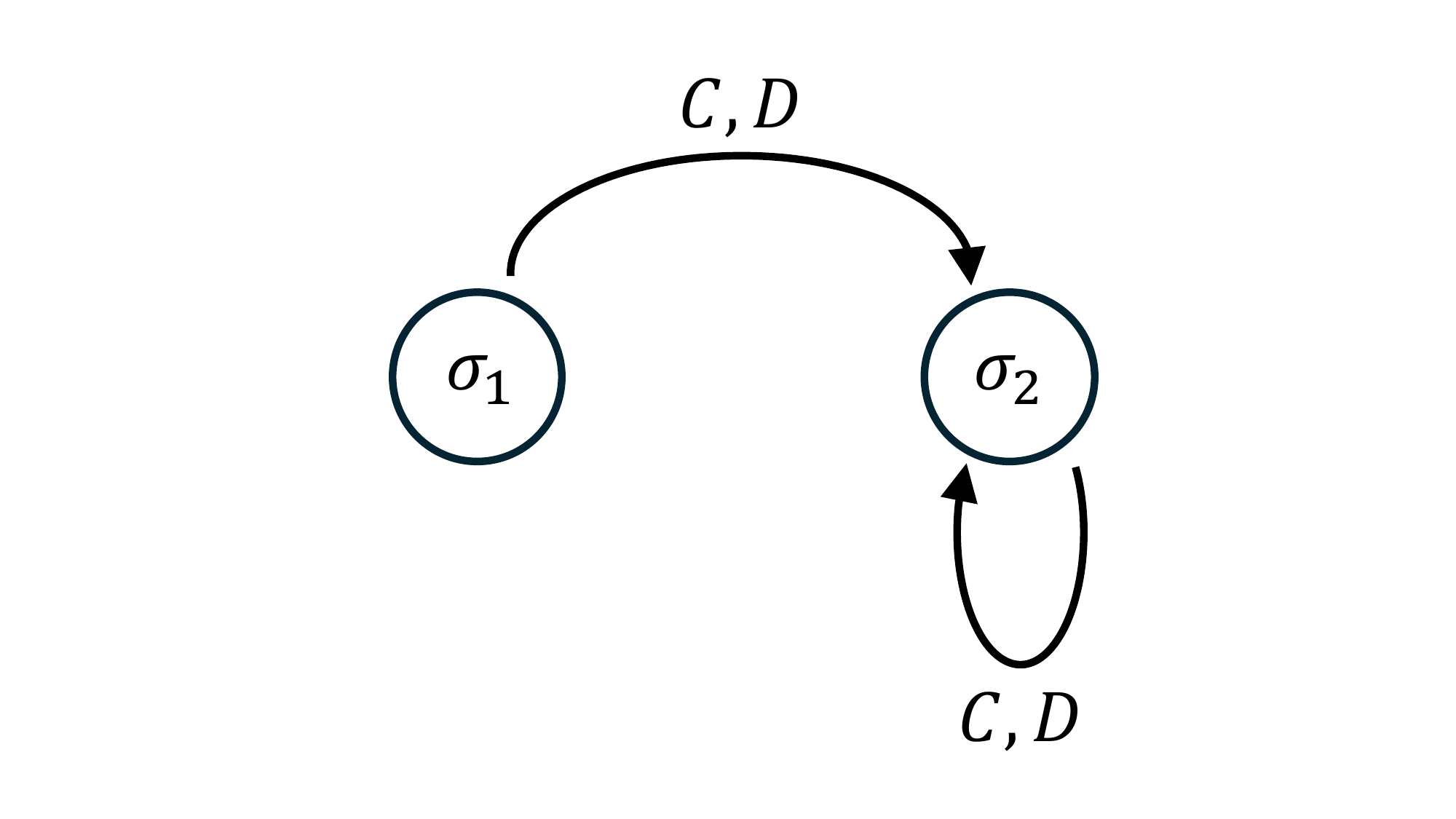}
\end{center}
\caption{A finite automaton in mode $\overline{a}_j$.}
\label{fig:SPD_ov}
\end{figure}

\begin{figure}[tbp]
\begin{center}
\includegraphics[clip, width=8.0cm]{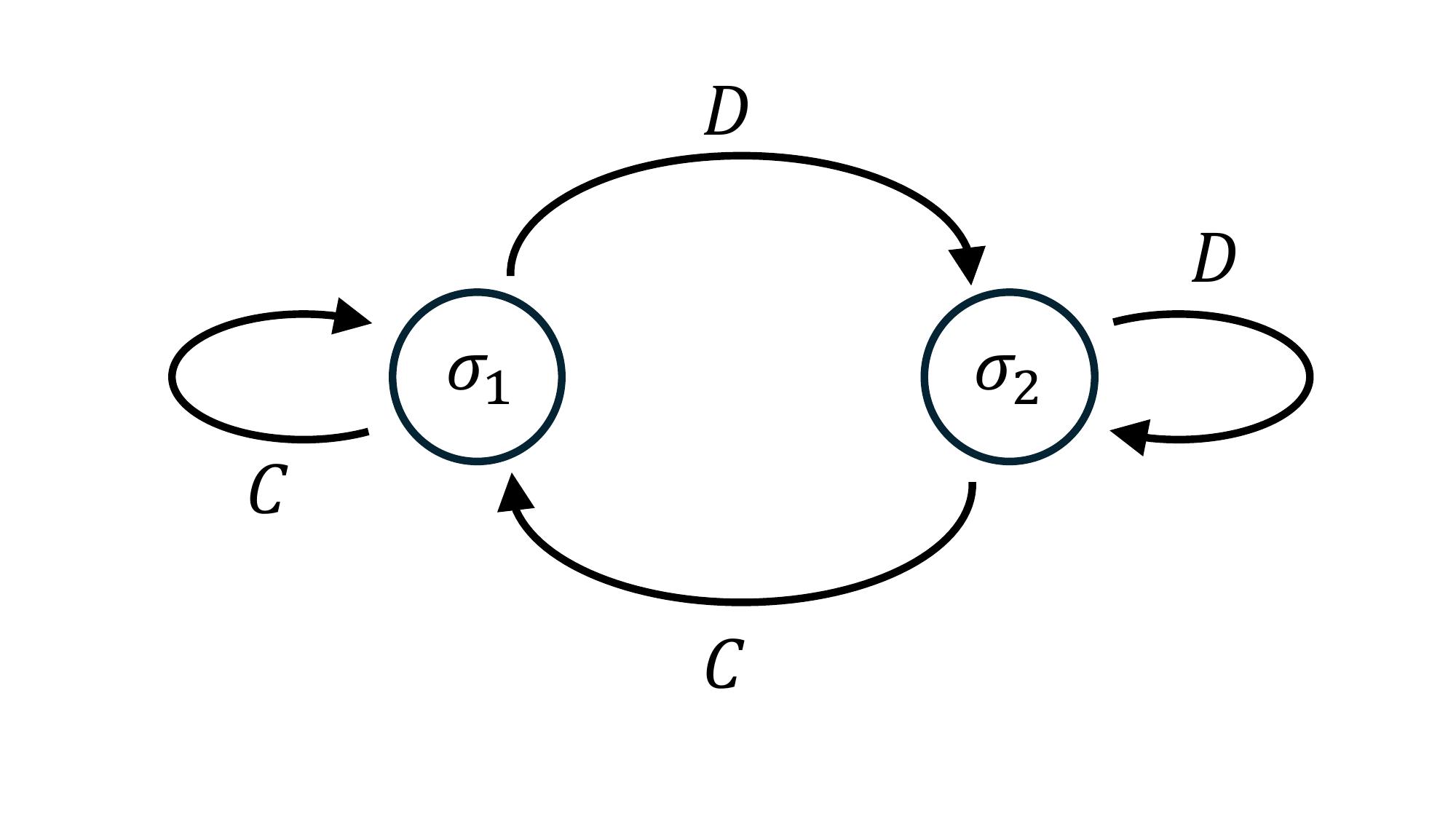}
\end{center}
\caption{A finite automaton in mode $\underline{a}_j$.}
\label{fig:SPD_und}
\end{figure}

For this model, we consider a feedback strategy of player $1$ about $B=s_1-s_2$ (fair feedback strategy).
According to the definitions (\ref{eq:aj_ov}) and (\ref{eq:aj_und}), we obtain $\overline{a}_j\left( \sigma \right)=D$ and $\underline{a}_j\left( \sigma \right)=C$ for all $\sigma$.
Therefore, this model satisfies Assumption \ref{asp:singleton}, and we can apply Theorem \ref{thm:existence}.
Because $\left| \Sigma \right|=2$, the length of a cycle is at most two.
Finite automaton representation of modes $\overline{a}_j$ and $\underline{a}_j$ is displayed in Figures \ref{fig:SPD_ov} and \ref{fig:SPD_und}, respectively.
In mode $\overline{a}_j$, possible cycles are
\begin{align}
 \left( D, C, \sigma_2 \right) &\rightarrow \cdots \\
 \left( D, D, \sigma_2 \right) &\rightarrow \cdots.
\end{align}
We call these cycles $C_1$ and $C_2$, respectively.
In cycle $C_1$, total $B$ is
\begin{align}
 \sum_{t=1}^{T_{C_1}} B\left( \overline{a}_j\left( \sigma^{(t)} \right), a_{-j}^{(t)}, \sigma^{(t)} \right) &= T_2 - S_2 \geq 0.
\end{align}
In cycle $C_2$, total $B$ is
\begin{align}
 \sum_{t=1}^{T_{C_2}} B\left( \overline{a}_j\left( \sigma^{(t)} \right), a_{-j}^{(t)}, \sigma^{(t)} \right) &= 0 \geq 0.
\end{align}
In mode $\underline{a}_j$, possible cycles are
\begin{align}
 \left( C, C, \sigma_1 \right) &\rightarrow \cdots \\
 \left( C, D, \sigma_2 \right) &\rightarrow \cdots \\
 \left( C, D, \sigma_1 \right) &\rightarrow \left( C, C, \sigma_2 \right) \rightarrow \cdots
\end{align}
We call these cycles $C_3$, $C_4$, and $C_5$, respectively.
In cycle $C_3$, total $B$ is
\begin{align}
 \sum_{t=1}^{T_{C_3}} B\left( \underline{a}_j\left( \sigma^{(t)} \right), a_{-j}^{(t)}, \sigma^{(t)} \right) &= 0 \leq 0.
\end{align}
In cycle $C_4$, total $B$ is
\begin{align}
 \sum_{t=1}^{T_{C_4}} B\left( \underline{a}_j\left( \sigma^{(t)} \right), a_{-j}^{(t)}, \sigma^{(t)} \right) &= S_2 - T_2 \leq 0.
\end{align}
In cycle $C_5$, total $B$ is
\begin{align}
 \sum_{t=1}^{T_{C_5}} B\left( \underline{a}_j\left( \sigma^{(t)} \right), a_{-j}^{(t)}, \sigma^{(t)} \right) &= S_1 - T_1 + 0 \leq 0.
\end{align}
Therefore, according to Theorem \ref{thm:existence}, the feedback strategy of player $1$ about $B=s_1-s_2$ is autocratic.
It should be noted that this feedback strategy is essentially the same as the strategy of cumulative reciprocity \cite{LZLet2022}, whereas we consider a stochastic game.

\subsection{Periodic prisoner's dilemma}
\label{subsec:PPD}
As another example, we consider a periodic prisoner's dilemma game \cite{MMHet2025,NakUed2027}.
In this game, $\mathcal{N}= \{ 1, 2 \}$, $\Sigma= \{ \sigma_1, \sigma_2 \}$, $A_j= \left\{ C, D \right\}$ $(j=1, 2)$, and the payoffs $\left\{ s_j \right\}_{j\in \mathcal{N}}$ are given as in Tables \ref{table:PD1} and \ref{table:PD2}, where $\bar{T}>\bar{R}>\bar{P}>\bar{S}$ and $2\bar{R}>\bar{T}+\bar{S}$ are assumed.
The transition function $f_\mathrm{E}$ is periodic:
\begin{align}
 f_\mathrm{E}\left( \bm{a}, \sigma_1 \right) &= \sigma_2 \quad \left( \forall \bm{a} \right) \nonumber \\
 f_\mathrm{E}\left( \bm{a}, \sigma_2 \right) &= \sigma_1 \quad \left( \forall \bm{a} \right).
\end{align}
For this game, the roles of two states $\sigma_1$ and $\sigma_2$ are symmetric.
\begin{table}[tb]
  \centering
  \caption{Payoffs in state $\sigma_1$.}
  \begin{tabular}{|c|cc|} \hline
    & $C$ & $D$ \\ \hline
   $C$ & $\bar{R} + \delta_R, \bar{R} - \delta_R$ & $\bar{S} + \delta_S, \bar{T} - \delta_T$ \\
   $D$ & $\bar{T} + \delta_T, \bar{S} - \delta_S$ & $\bar{P} + \delta_P, \bar{P} - \delta_P$ \\ \hline
  \end{tabular}
  \label{table:PD1}
  
  \centering
  \caption{Payoffs in state $\sigma_2$.}
  \begin{tabular}{|c|cc|} \hline
    & $C$ & $D$ \\ \hline
   $C$ & $\bar{R} - \delta_R, \bar{R} + \delta_R$ & $\bar{S} - \delta_S, \bar{T} + \delta_T$ \\
   $D$ & $\bar{T} - \delta_T, \bar{S} + \delta_S$ & $\bar{P} - \delta_P, \bar{P} + \delta_P$ \\ \hline
  \end{tabular}
  \label{table:PD2}
\end{table}

For this model, we again consider a feedback strategy of player $1$ about $B=s_1-s_2$ (fair feedback strategy).
According to Theorem \ref{thm:existence_periodic}, the existence condition of the feedback autocratic strategy is
\begin{align}
 \sum_{t=1}^{2} \max_{a_1} \min_{a_{2}} B\left( a_1, a_{2}, \sigma_t \right) &\geq 0 \nonumber \\
 \sum_{t=1}^{2} \min_{a_1} \max_{a_{2}} B\left( a_1, a_{2}, \sigma_t \right) &\leq 0. 
\end{align}
Concretely, this condition is written as
\begin{align}
 &\max\left\{ \min\left\{ 2\delta_R, \bar{S}-\bar{T}+\delta_S+\delta_T \right\}, \min\left\{ \bar{T}-\bar{S}+\delta_S+\delta_T, 2\delta_P \right\} \right\} \nonumber \\
 & \quad +\max\left\{ \min\left\{ -2\delta_R, \bar{S}-\bar{T}-\delta_S-\delta_T \right\}, \min\left\{ \bar{T}-\bar{S}-\delta_S-\delta_T, -2\delta_P \right\} \right\} \geq 0 \nonumber \\
 &\min\left\{ \max\left\{ 2\delta_R, \bar{S}-\bar{T}+\delta_S+\delta_T \right\}, \max\left\{ \bar{T}-\bar{S}+\delta_S+\delta_T, 2\delta_P \right\} \right\} \nonumber \\
 & \quad +\min\left\{ \max\left\{ -2\delta_R, \bar{S}-\bar{T}-\delta_S-\delta_T \right\}, \max\left\{ \bar{T}-\bar{S}-\delta_S-\delta_T, -2\delta_P \right\} \right\} \leq 0.
\end{align}
This condition is further rewritten as
\begin{align}
 &\max\left\{ \min\left\{ 2\delta_R, \bar{S}-\bar{T}+\delta_S+\delta_T \right\}, \min\left\{ \bar{T}-\bar{S}+\delta_S+\delta_T, 2\delta_P \right\} \right\} \nonumber \\
 & \quad -\min\left\{ \max\left\{ 2\delta_R, \bar{T}-\bar{S}+\delta_S+\delta_T \right\}, \max\left\{ \bar{S}-\bar{T}+\delta_S+\delta_T, 2\delta_P \right\} \right\} \geq 0 \nonumber \\
 &\min\left\{ \max\left\{ 2\delta_R, \bar{S}-\bar{T}+\delta_S+\delta_T \right\}, \max\left\{ \bar{T}-\bar{S}+\delta_S+\delta_T, 2\delta_P \right\} \right\} \nonumber \\
 & \quad -\max\left\{ \min\left\{ 2\delta_R, \bar{T}-\bar{S}+\delta_S+\delta_T \right\}, \min\left\{ \bar{S}-\bar{T}+\delta_S+\delta_T, 2\delta_P \right\} \right\} \leq 0.
 \label{eq:existence_per_fair}
\end{align}
We introduce the notations $D_1:= \bar{S}-\bar{T}+\delta_S+\delta_T$ and $D_2:= \bar{T}-\bar{S}+\delta_S+\delta_T$ with $D_2>D_1$.
We consider 12 cases separately.
\begin{enumerate}
\item $D_1 < D_2 \leq 2\delta_R \leq 2\delta_P$\\
For this case, the first inequality in Eq. (\ref{eq:existence_per_fair}) is
\begin{align}
 \max\left\{ D_1, D_2 \right\} - \min\left\{ 2\delta_R, 2\delta_P \right\} &\geq 0,
\end{align}
that is,
\begin{align}
 D_2 - 2\delta_R &\geq 0.
\end{align}
This is satisfied only for the case $D_2 = 2\delta_R$.
The second inequality in Eq. (\ref{eq:existence_per_fair}) is
\begin{align}
 \min\left\{ 2\delta_R, 2\delta_P \right\} - \max\left\{ D_2, D_1 \right\} &\leq 0,
\end{align}
that is,
\begin{align}
 2\delta_R - D_2 &\leq 0.
\end{align}
This is satisfied only for the case $D_2 = 2\delta_R$.

\item $D_1 \leq 2\delta_R < D_2 \leq 2\delta_P$\\
For this case, the first inequality in Eq. (\ref{eq:existence_per_fair}) is
\begin{align}
 \max\left\{ D_1, D_2 \right\} - \min\left\{ D_2, 2\delta_P \right\} &\geq 0,
\end{align}
that is,
\begin{align}
 D_2 - D_2 &\geq 0.
\end{align}
This is satisfied.
The second inequality in Eq. (\ref{eq:existence_per_fair}) is
\begin{align}
 \min\left\{ 2\delta_R, 2\delta_P \right\} - \max\left\{ 2\delta_R, D_1 \right\} &\leq 0,
\end{align}
that is,
\begin{align}
 2\delta_R - 2\delta_R &\leq 0.
\end{align}
This is satisfied.

\item $D_1 \leq 2\delta_R \leq 2\delta_P < D_2$\\
For this case, the first inequality in Eq. (\ref{eq:existence_per_fair}) is
\begin{align}
 \max\left\{ D_1, 2\delta_P \right\} - \min\left\{ D_2, 2\delta_P \right\} &\geq 0,
\end{align}
that is,
\begin{align}
 2\delta_P - 2\delta_P &\geq 0.
\end{align}
This is satisfied.
The second inequality in Eq. (\ref{eq:existence_per_fair}) is
\begin{align}
 \min\left\{ 2\delta_R, D_2 \right\} - \max\left\{ 2\delta_R, D_1 \right\} &\leq 0,
\end{align}
that is,
\begin{align}
 2\delta_R - 2\delta_R &\leq 0.
\end{align}
This is satisfied.

\item $2\delta_R \leq D_1 < D_2 \leq 2\delta_P$\\
For this case, the first inequality in Eq. (\ref{eq:existence_per_fair}) is
\begin{align}
 \max\left\{ 2\delta_R, D_2 \right\} - \min\left\{ D_2, 2\delta_P \right\} &\geq 0,
\end{align}
that is,
\begin{align}
 D_2 - D_2 &\geq 0.
\end{align}
This is satisfied.
The second inequality in Eq. (\ref{eq:existence_per_fair}) is
\begin{align}
 \min\left\{ D_1, 2\delta_P \right\} - \max\left\{ 2\delta_R, D_1 \right\} &\leq 0,
\end{align}
that is,
\begin{align}
 D_1 - D_1 &\leq 0.
\end{align}
This is satisfied.

\item $2\delta_R \leq D_1 \leq 2\delta_P < D_2$\\
For this case, the first inequality in Eq. (\ref{eq:existence_per_fair}) is
\begin{align}
 \max\left\{ 2\delta_R, 2\delta_P \right\} - \min\left\{ D_2, 2\delta_P \right\} &\geq 0,
\end{align}
that is,
\begin{align}
 2\delta_P - 2\delta_P &\geq 0.
\end{align}
This is satisfied.
The second inequality in Eq. (\ref{eq:existence_per_fair}) is
\begin{align}
 \min\left\{ D_1, D_2 \right\} - \max\left\{ 2\delta_R, D_1 \right\} &\leq 0,
\end{align}
that is,
\begin{align}
 D_1 - D_1 &\leq 0.
\end{align}
This is satisfied.

\item $2\delta_R \leq 2\delta_P \leq D_1 < D_2$\\
For this case, the first inequality in Eq. (\ref{eq:existence_per_fair}) is
\begin{align}
 \max\left\{ 2\delta_R, 2\delta_P \right\} - \min\left\{ D_2, D_1 \right\} &\geq 0,
\end{align}
that is,
\begin{align}
 2\delta_P - D_1 &\geq 0.
\end{align}
This is satisfied only for the case $D_1 = 2\delta_P$.
The second inequality in Eq. (\ref{eq:existence_per_fair}) is
\begin{align}
 \min\left\{ D_1, D_2 \right\} - \max\left\{ 2\delta_R, 2\delta_P \right\} &\leq 0,
\end{align}
that is,
\begin{align}
 D_1 - 2\delta_P &\leq 0.
\end{align}
This is satisfied only for the case $D_1 = 2\delta_P$.

\item $D_1 < D_2 \leq 2\delta_P \leq 2\delta_R$\\
This case corresponds to case 1 with replacement $\delta_R \leftrightarrow \delta_P$.
However, two inequalities in Eq. (\ref{eq:existence_per_fair}) are also exchanged by the replacement $\delta_R \leftrightarrow \delta_P$.
Therefore, this case is reduced to case 1.

\item $D_1 \leq 2\delta_P < D_2 \leq 2\delta_R$\\
This case corresponds to case 2 with replacement $\delta_R \leftrightarrow \delta_P$.
However, two inequalities in Eq. (\ref{eq:existence_per_fair}) are also exchanged by the replacement $\delta_R \leftrightarrow \delta_P$.
Therefore, this case is reduced to case 2.

\item $D_1 \leq 2\delta_P \leq 2\delta_R < D_2$\\
This case corresponds to case 3 with replacement $\delta_R \leftrightarrow \delta_P$.
However, two inequalities in Eq. (\ref{eq:existence_per_fair}) are also exchanged by the replacement $\delta_R \leftrightarrow \delta_P$.
Therefore, this case is reduced to case 3.

\item $2\delta_P \leq D_1 < D_2 \leq 2\delta_R$\\
This case corresponds to case 4 with replacement $\delta_R \leftrightarrow \delta_P$.
However, two inequalities in Eq. (\ref{eq:existence_per_fair}) are also exchanged by the replacement $\delta_R \leftrightarrow \delta_P$.
Therefore, this case is reduced to case 4.

\item $2\delta_P \leq D_1 \leq 2\delta_R < D_2$\\
This case corresponds to case 5 with replacement $\delta_R \leftrightarrow \delta_P$.
However, two inequalities in Eq. (\ref{eq:existence_per_fair}) are also exchanged by the replacement $\delta_R \leftrightarrow \delta_P$.
Therefore, this case is reduced to case 5.

\item $2\delta_P \leq 2\delta_R \leq D_1 < D_2$\\
This case corresponds to case 6 with replacement $\delta_R \leftrightarrow \delta_P$.
However, two inequalities in Eq. (\ref{eq:existence_per_fair}) are also exchanged by the replacement $\delta_R \leftrightarrow \delta_P$.
Therefore, this case is reduced to case 6.
\end{enumerate}
From the case analysis, we find that we can construct the feedback autocratic strategy controlling $B=s_1-s_2$ for cases 1 with $D_2 = 2\delta_R$, 2, 3, 4, 5, 6 with $D_1 = 2\delta_P$, 7 with $D_2 = 2\delta_P$, 8, 9, 10, 11, and 12 with $D_1 = 2\delta_R$.

It should be noted that memory-one zero-determinant strategies controlling $B=s_1-s_2$ exist only for cases 3, 4, 9, and 10 \cite{NakUed2027}.
Therefore, by considering feedback strategies, we can equalize the payoffs of two players in broader situations.

In order to check the validity, we investigate several cases in detail.
In case 1 with $D_2 < 2\delta_R$, $\overline{a}_j\left( \sigma_1 \right)=D$, $\overline{a}_j\left( \sigma_2 \right)=C$, $\underline{a}_j\left( \sigma_1 \right)=C$, and $\underline{a}_j\left( \sigma_2 \right)=C$.
Therefore, if we consider the cycle
\begin{align}
 \left( D, C, \sigma_1 \right) &\rightarrow \left( C, C, \sigma_2 \right) \rightarrow \cdots,
\end{align}
total $B$ is
\begin{align}
 \sum_{t=1}^{2} B\left( \overline{a}_j\left( \sigma_t \right), a_{2}^{(t)}, \sigma_t \right) &= D_2 + \left( -2\delta_R \right) < 0.
\end{align}
In addition, this means that at least either $D_2$ or $-2\delta_R$ is negative.
Therefore, if the feedback strategy starts with negative $B$, this cycle can be infinitely repeated, and the feedback strategy cannot be autocratic.
In contrast, in case 2, $\overline{a}_j\left( \sigma_1 \right)=D$, $\overline{a}_j\left( \sigma_2 \right)=C$, $\underline{a}_j\left( \sigma_1 \right)=C$, and $\underline{a}_j\left( \sigma_2 \right)=C$.
For this case, all possible cycles in mode $\overline{a}_j$ are
\begin{align}
 \left( D, C, \sigma_1 \right) &\rightarrow \left( C, C, \sigma_2 \right) \rightarrow \cdots \\
 \left( D, C, \sigma_1 \right) &\rightarrow \left( C, D, \sigma_2 \right) \rightarrow \cdots \\
 \left( D, D, \sigma_1 \right) &\rightarrow \left( C, C, \sigma_2 \right) \rightarrow \cdots \\
 \left( D, D, \sigma_1 \right) &\rightarrow \left( C, D, \sigma_2 \right) \rightarrow \cdots.
\end{align}
For each cycle, total $B$ is
\begin{align}
 D_2 + \left( -2\delta_R \right) &\geq 0 \\
 D_2 + \left( - D_2 \right) &\geq 0 \\
 2\delta_P + \left( -2\delta_R \right) &\geq 0 \\
 2\delta_P + \left( -D_2 \right) &\geq 0,
\end{align}
respectively.
Furthermore, all possible cycles in mode $\underline{a}_j$ are
\begin{align}
 \left( C, C, \sigma_1 \right) &\rightarrow \left( C, C, \sigma_2 \right) \rightarrow \cdots \\
 \left( C, C, \sigma_1 \right) &\rightarrow \left( C, D, \sigma_2 \right) \rightarrow \cdots \\
 \left( C, D, \sigma_1 \right) &\rightarrow \left( C, C, \sigma_2 \right) \rightarrow \cdots \\
 \left( C, D, \sigma_1 \right) &\rightarrow \left( C, D, \sigma_2 \right) \rightarrow \cdots.
\end{align}
For each cycle, total $B$ is
\begin{align}
 2\delta_R + \left( -2\delta_R \right) &\leq 0 \\
 2\delta_R + \left( - D_2 \right) &\leq 0 \\
 D_1 + \left( -2\delta_R \right) &\leq 0 \\
 D_1 + \left( -D_2 \right) &\leq 0,
\end{align}
respectively.
Therefore, the feedback strategy is autocratic.

\section{Concluding remarks}
\label{sec:remarks}
In this paper, we introduced the feedback strategies, and investigated the condition in which they become autocratic.
Furthermore, we showed that the condition in periodic games is reduced to the existence condition of autocratic strategies in multichannel games.
We demonstrated the main results in the stochastic prisoner's dilemma game and the periodic prisoner's dilemma game.

Before ending this paper, we make four remarks.
The first remark is related to exploration of cycles.
In Theorem \ref{thm:existence}, we have to check the conditions (\ref{eq:condition_ov}) and (\ref{eq:condition_und}) for all possible cycles.
However, according to the structure of a graph, the number of cycles can increase exponentially as the number of vertices (environmental states) increases.
Therefore, checking the conditions (\ref{eq:condition_ov}) and (\ref{eq:condition_und}) for all possible cycles can be computationally hard.
It will be useful if we can obtain some alternative conditions which are equivalent to the conditions (\ref{eq:condition_ov}) and (\ref{eq:condition_und}).

The second remark is related to choice of the definitions (\ref{eq:aj_ov}) and (\ref{eq:aj_und}).
As we remarked in Section \ref{subsec:existence}, choice of $\overline{a}_j$ and $\underline{a}_j$ has influence on the performance of feedback strategies.
We do not know whether the definitions (\ref{eq:aj_ov}) and (\ref{eq:aj_und}) are best or not.
For periodic games, according to Theorem \ref{thm:existence_periodic}, these definitions are best, since if any other feedback strategies are autocratic, the condition (\ref{eq:existence_multichannel}) is satisfied and the feedback strategies with the definitions (\ref{eq:aj_ov}) and (\ref{eq:aj_und}) are also autocratic.
However, we do not know whether this property holds even in general stochastic games, because cycles depend on the definitions of $\overline{a}_j$ and $\underline{a}_j$.
Finding the best definitions of $\overline{a}_j$ and $\underline{a}_j$ is an important future problem.

The third remark is related to our assumptions.
In this paper, we made two assumptions (Assumptions \ref{asp:deterministic} and \ref{asp:singleton}).
Without these assumptions, the dynamics of the system becomes stochastic.
Therefore, the analysis will be more complicated.
Investigation of the condition for the existence of feedback autocratic strategies in general stochastic cases remains to be solved.

Finally, we remark on the relation to zero-determinant strategies.
Zero-determinant strategies, which also unilaterally control payoffs in repeated games, are memory-one strategies, which use only the information of an action profile in the previous round.
In contrast, the feedback strategies (\ref{eq:feedback}) are infinite-memory strategies, because they remember total $B$ in all previous rounds.
In the context of the equalizer strategies in the repeated prisoner's dilemma game, it was discussed that such short memory is characteristic to zero-determinant strategies \cite{BNS1997}.
In fact, in repeated games, the existence condition of zero-determinant strategies is equivalent to that of feedback autocratic strategies \cite{Ued2025}.
However, in stochastic games, how to define zero-determinant strategies as short-memory strategies is not clear, since the state changes as time evolves.
As far as we define zero-determinant strategies as memory-one strategies, there is a discrepancy between the existence condition of fair zero-determinant strategies \cite{NakUed2027} and that of fair feedback autocratic strategies; see Section \ref{subsec:PPD}.
Defining zero-determinant strategies which are equivalent to the feedback autocratic strategies in stochastic games is an open problem.

\section*{Acknowledgement}
This study was supported by JSPS KAKENHI Grant Number JP26K21335, and JST PRESTO, Japan, Grant Number JPMJPR26K1.

\appendix
\section{Lemma}
\label{app:lemma}

\begin{lemma}
\label{lem:cycle}
Suppose that the sequence $\{ b_t \}_{t=1}^\infty$ is periodic with period $T$.
If $\sum_{t=1}^T b_t<0$, then by choosing $t_0\in [1, T]$ appropriately the sequence satisfies
\begin{align}
 \sum_{t=t_0+1}^{t_0+\tau} b_t &< 0 \quad (\forall \tau\in [1, T]).
\end{align}
\end{lemma}

\begin{proof}
We define the sum
\begin{align}
 S_t &:= \sum_{t^\prime=1}^t b_{t^\prime}.
\end{align}
We also define
\begin{align}
 t^* &:= \arg \max_{1\leq t \leq T} S_t.
 \label{eq:cycle_t0}
\end{align}
If there are several $t^*$ which satisfy Eq. (\ref{eq:cycle_t0}), we choose the largest $t^*$.
It should be noted that
\begin{align}
 \sum_{t=t^*+1}^{t^*+\tau} b_t &= S_{t^*+\tau} - S_{t^*} \quad (\forall \tau\in [1, T]),
\end{align}
and the assumption is rewritten as $S_T<0$.

If $t^*+\tau \leq T$, by the definition of $t^*$, we obtain
\begin{align}
 \sum_{t=t^*+1}^{t^*+\tau} b_t &= S_{t^*+\tau} - S_{t^*} <0,
\end{align}
where we have used the above rule of tie-breaking.

If $T< t^*+\tau \leq t^*+T$, by periodicity and the definition of $t^*$, we obtain
\begin{align}
 \sum_{t=t^*+1}^{t^*+\tau} b_t &= S_{T} + S_{t^*+\tau-T} - S_{t^*} \leq S_{T} < 0.
\end{align}

Therefore, we can choose $t_0=t^*$.
\end{proof}

\section*{References}

\bibliography{ZDS}

\end{document}